\documentclass[12pt]{article}
\usepackage[latin1]{inputenc}
\usepackage{graphicx,epsfig}
\usepackage{amsmath}
\usepackage{amsfonts}
\usepackage{amssymb}
\usepackage{amsthm}
\usepackage{bm}
\usepackage{slashed}
\def\be{\begin{equation}}
\def\ee{\end{equation}}
\def\ba{\begin{eqnarray}}
\def\ea{\end{eqnarray}}
\def\beq{\begin{equation}}
\def\eeq{\end{equation}}
\def\bea{\begin{eqnarray}}
\def\eea{\end{eqnarray}}

\begin{document}

\thispagestyle{empty}

\begin{center}

{\Large {\bf CASIMIR ENERGY CORRECTIONS BY  LIGHT-CONE FLUCTUATIONS  }} \\
\vspace{1.0cm}
{\large {\bf E. Arias}\footnote{e-mail: enrike@cbpf.br},  {\bf J.
G. Due\~nas}\footnote{e-mail: jgduenas@cbpf.br},{\bf N. F. Svaiter}\footnote{e-mail: nfuxsvai@cbpf.br}}   \\
\vspace{0.2cm}
{\it  Centro Brasileiro de Pesquisas F\'{\i}sicas,  Rua Dr. Xavier Sigaud 150, 22290-180, Rio de Janeiro, RJ, Brazil} \\
\vspace{0.2cm}
{\large {\bf C. H. G. Bessa}\footnote{e-mail: carlos@cosmos.phy.tufts.edu}}\\
{\it Institute of Cosmology, Department of Physics and Astronomy\\Tufts University,
Medford, Massachusetts 02155, USA}\\
\vspace{0.2cm}
{\large {\bf G. Menezes}\footnote{e-mail: gsm@ift.unesp.br}}\\
{\it Instituto de F\'\i sica Te\'orica, Universidade Estadual Paulista\\
Rua Dr. Bento Teobaldo Ferraz 271 - Bloco II, 01140-070  S\~ao Paulo, SP, Brazil}

\end{center}


\begin{abstract}
We study the effects of light-cone fluctuations on the renormalized zero-point energy associated with a 
free massless scalar field in the presence of boundaries. In order to simulate light-cone 
fluctuations we introduce a space-time dependent random coefficient in the Klein-Gordon operator. 
We assume that the field  is defined in a domain with one confined direction. For simplicity, 
we choose the symmetric case of two parallel plates separated by a distance $a$. 
The correction to the renormalized vacuum energy density between the plates goes as 
$1/a^{8}$ instead of the usual $1/a^{4}$ dependence for the free case. 
In turn we also show that light-cone fluctuations break down the vacuum pressure homogeneity between the plates. 
\end{abstract}


\newpage




    \section{Introduction}

One consequence of the quantization of any classical field is the occurrence of a divergent zero-point energy. 
Casimir predicted that uncharged, parallel, perfectly conducting plates should attract each other with a finite 
force ~\cite{casimir}. This force can be interpreted as being the distortion of the zero-point energy of the 
electromagnetic field due to the presence of the plates. There are several reviews discussing such effect. 
See for example ~\cite{plunien,mostepanenko,milton,Bordag} and more recently~\cite{Bordag2,qftex09,milton11,lectphys}.

In a global approach, the Casimir energy can be obtained adopting the following prescription: the eigenfrequencies 
of the mode solutions to the classical wave equation with given boundary conditions are found; the divergent zero-point
energy of the quantized field is regularized by the introduction of a  cut-off and then renormalized using auxiliary
configurations which are added and subtracted. This procedure is justified in the absence of gravity, since in 
non-gravitational physics only energy differences are measurable. 

When gravity is taken into account, this procedure is no longer acceptable. See for example~\cite{Isham2,davies}. 
In this situation, a reasonable approximation is to treat the gravitational field as a classical background field. 
In this semiclassical approximation important effects were predicted, as for instance particle creation in cosmological 
models~\cite{parker} and black hole evaporation~\cite{hawking, hawking2}. 

There are different proposals to go beyond the semiclassical approximation for general relativity. Let us discuss briefly
two different approaches which form the basis of the model presented below. One is the stochastic gravity program, where
the so-called Einstein-Langevin equation enables one to find the dynamics of metric fluctuations generated by the
fluctuations of the stress tensor associated with quantum fields~\cite{bei}. On the other hand, one of the primary
consequences of assuming that gravity obeys quantum-mechanical laws is that the structure of space-time is bound to undergo
quantum fluctuations. The approach of a classical background with a light-cone defined in each of its points must be modified 
if one introduce quantum mechanical laws in the framework of general relativity. For instance, Ford and collaborators 
discussed different situations where a bath of gravitons induces light-cone fluctuations \cite{ford11,fn2,fn2a,fn2b,fn2c,fn2d,fn2e}.
The wave propagation in stochastic space-time in a curved background was analyzed in Ref. \cite{shiok}. Whereas the origins of the 
stochasticity is not discussed, it was shown that the metric fluctuations can be represented as fluctuations in the optical
index of the wave.

In other hand a lot of activity exploring the analogy between quantized sound waves in fluids and quantum fields in curved space
has been developed~\cite{unruh,analog}. In this framework, quantized acoustic perturbations in the presence of boundaries produces
the phononic Casimir effect \cite{nami2,nami3,casimirbec}. Before proceed, we would like to call reader's attention for two papers
related to the present discussion. The first one is Ref. \cite{di4} where the thermal Casimir effect between random layered
dielectrics was studied. For the Casimir effect in dielectric, see for example \cite{di1,di2,di3}.
The second one is Ref. \cite{time} where the scalar wave equation with a time-dependent random dielectric constant was discussed.
In the following, we are using the ideas developed in the above two mentioned papers to discuss the Casimir effect in the presence
of light-cone fluctuations. Having in mind the aforementioned analogy between quantized sound waves in random fluids and quantum
fields in a space-time with metric fluctuations, one could regard as ``sound-cone fluctuations" the random fluctuations treated in this paper. 

The effect of light-cone fluctuations over quantum fields was considered in different situations, see Refs.~\cite{krein1,arias,arias2,bessa,Fordvitorio}.
In these papers  the  light-cone fluctuations were described by random differential equations. More recently the corrections due to 
light-cone fluctuations for the renormalized vacuum expectation value of the stress-energy-momentum tensor associated with a scalar
field were considered~\cite{vitorio}. The scalar field was defined in a $(d+1)$-dimensional flat space-time with non-trivial topology. 
For papers discussing quantum field theory in a nonsimply connected space-time see for example~\cite{non-topology}. 

Here we intend to extend such analysis to the case for a free massless scalar field in the presence of boundaries. We investigate the effects
of light-cone fluctuations on the renormalized vacuum expectation value of the stress tensor of a quantum field confined between plane 
boundaries. We assume that such effects on quantum fields can be described using random differential equations~\cite{pe13}. For the case of 
a scalar field we consider a random Klein-Gordon equation. In order to calculate the renormalized vacuum energy of the system, we employ 
the point-splitting method of Green's functions~\cite{Green's,Green's2,Green's3,Green's4,Fulling}. Our result generalizes to some extent 
the ones obtained by Edery~\cite{edery} where the Casimir effect in a relativistic perfect fluid was investigated.

The organization of the paper is as follows. In Section~\ref{sec2} we present the local method of Green's functions to calculate the 
vacuum energy. In Section~\ref{sec3} we discuss modifications in the wave equation for a free quantum scalar field due to randomness. 
In addition, we calculate within a perturbative approach the Green's function of the scalar field in the presence of boundaries.
In this framework, we compute the corrections to the renormalized stress-energy-momentum tensor of the system. We show that the correction
to the vacuum energy density between the plates goes as $1/a^{8}$ instead of the usual $1/a^{4}$ dependence for the free case. 
In Section~\ref{sec3} details of these local contributions are presented. Due to light-cone fluctuations the vacuum pressure acquires a dependence 
on the distance to the plates. Conclusions are given in Section~\ref{sec5}. The paper includes Appendices containing details of lengthy derivations.
Throughout the paper we employ units with $\hbar=c=1$.


 \section{The renormalized stress tensor of a scalar field confined between plane boundaries}\label{sec2}

The aim of this Section is to use the point-splitting method to obtain the renormalized stress-energy momentum tensor associated with a minimally
coupled massless scalar field in the presence of parallel plates. We do not consider randomness in the system at this point.The Lagrangian density of 
this system is given by
    \begin{equation}
    {\cal L}=\frac{1}{2}\partial_\mu\varphi\partial^\mu\varphi,
    \end{equation}
and the associated stress-energy-momentum tensor (stress tensor for short) is given by
    \begin{equation}
    T^{\mu\nu}(x)=\partial^\mu\varphi\partial^\nu\varphi-g^{\mu\nu}{\cal L},
    \end{equation}
Taking the vacuum expectation value of the stress tensor and defining the causal Green's function as $iG(x,x')=\langle0|T[\varphi(x)\varphi(x')]|0\rangle$,
where $T$ is the time-order product, one has
    \begin{equation}
    \langle0|T^{\mu\nu}(x)|0\rangle=\lim_{x'\rightarrow x}
    \left(\partial^\mu\partial'^\nu-\frac{1}{2}g^{\mu\nu}\partial_\alpha\partial'^\alpha\right)iG(x,x').
    \end{equation}
We are interested in the case where the field is confined between two parallel plates. We will denote the $(d+1)$ space-time coordinates as 
$x^\mu=(t,{\bf x})$ and the $d$-spatial coordinates as ${\bf x}=({\bf x}_\bot,z)$, where ${\bf x_\bot}$ are the unconfined coordinates and $z$ is 
the spatial coordinate between the plates. Assuming that the plates are located at $z=0$ and $z=a$ and Dirichlet boundary conditions, we have
    \begin{eqnarray}
    (\partial_t^2-\nabla_\bot^2-\partial_z^2)\varphi(x)=0,&&\nonumber\\
    \varphi(x)|_{z=0}=\varphi(x)|_{z=a}=0.&&
    \end{eqnarray}
Therefore in the presence of boundaries the free Green's function satisfies
    \begin{eqnarray}
    (\partial_t^2-\nabla_\bot^2-\partial_z^2)G_0(x,x')=-\delta^{(d+1)}(x-x'),&&\nonumber\\
    G_0(x,x')|_{z=0}=G_0(x,x')|_{z=a}=0.&&
    \end{eqnarray}
One may employ a Fourier representation for the Green's function
    \begin{equation}
    G_0(x,x')=\int
    \frac{d\omega}{2\pi}\frac{d^{d-1}{\bf k}_\bot}{(2\pi)^{d-1}}\exp[-i\omega(t-t')+i{\bf k}_\bot({\bf x}_\bot-{\bf x}'_\bot)]
    {\cal G}_\lambda(z,z'),
    \label{G0}
    \end{equation}
 where the function ${\cal G}_\lambda(z,z')$ only depends on the confined coordinates and satisfies
    \begin{eqnarray}
    (\partial_z^2+\lambda^2){\cal G}_\lambda(z,z')=\delta^{(1)}(z-z'),&&\nonumber\\
    {\cal G}_\lambda(z,z')|_{z=0}={\cal G}_\lambda(z,z')|_{z=a}=0,&&
    \end{eqnarray}
 where $\lambda=\sqrt{\omega^2-{\bf k}_\bot^2}$. The solution to the above equation is given by
    \begin{equation}
    {\cal G}_\lambda(z,z')=\frac{2}{a}\sum_{n=1}^\infty\frac{\sin\left(\frac{n\pi z}{a}\right)\sin\left(\frac{n\pi z'}{a}\right)}
    {\lambda^2-\left(\frac{n\pi}{a}\right)^2}.
    \label{gl0}
    \end{equation}
 Using the point splitting method, the energy density associated with the scalar field is given by
    \begin{equation}
    \langle T^{00}(x)\rangle_{0}=\lim_{x'\rightarrow x}\frac{i}{2}\left(\partial_t\partial_{t'}+
    \partial_{\bf x_\bot}\partial_{\bf x'_\bot}+\partial_z\partial_{z'}\right)G_0(x,x'),
    \end{equation}
 where the subscript on the left-hand side of the above equation is to denote the situation without randomness. In this case we find that
    \begin{eqnarray}
    \langle T^{00}(x)\rangle_{0}&=&\frac{i}{a}\int\frac{d\omega}{2\pi}\frac{d^{d-1}{\bf k}_\bot}{(2\pi)^{d-1}}
    \sum_{n=1}^\infty\bigg[\frac{\omega^2+\mathbf{k}_\bot^2}{\omega^2-\overline{\omega}_n^2+i\epsilon}\sin^2(n\pi z/a)\nonumber\\
    &&+\frac{(n\pi/a)^2}{\omega^2-\overline{\omega}_n^2+i\epsilon}\cos^2(n\pi z/a)\bigg].
    \label{eqT00}
    \end{eqnarray}
 In the above equation we have defined $\overline{\omega}_n=\sqrt{\mathbf{k}_\bot^2+(n\pi/a)^2}$. From Eq.~(\ref{eqT00}) the energy density
 only depends on the confined coordinate $z$. A straightforward calculation yields
    \begin{equation}
    \langle T^{00}(z)\rangle_{0}=-\frac{1}{2a^{d+1}}\left(\frac{\pi}{4}\right)^{d/2}\Gamma\left(-\frac{d}{2}\right)
    \bigg[\zeta(-d)+(d-1)K(z)\bigg],
    \label{T00free}
    \end{equation}
 where $\zeta(s)$ is the usual Riemann zeta function and the function $K(z)$ is given by
    \begin{equation}
    K(z)=\sum_{n=1}^{\infty}n^d\cos\left(\frac{2n\pi z}{a}\right).
    \label{k12}    
    \end{equation}
The first term on the right hand side of Eq. (\ref{T00free}), which is global will be denoted by $U_0(a)$. This will be identified as the 
Casimir energy density for the free case. By using the reflection property of the Riemann zeta function 
	\begin{equation}
	\Gamma\left(\frac{z}{2}\right)\zeta(z)\pi^{-z/2} = \Gamma\left(\frac{1-z}{2}\right)\zeta(1-z)\pi^{(z-1)/2},
	\label{reflection}
	\end{equation} 
	one can write
	\begin{equation}
	 U_0(a)= - \frac{(4\pi)^{-(d+1)/2}}{a^{d+1}}\Gamma\left(\frac{d+1}{2}\right)\zeta(d+1).
	 \label{fernegy}
	\end{equation}

\begin{figure}[h]
 \centering
 \includegraphics[scale=0.62]{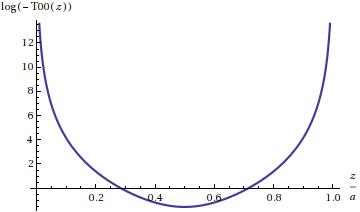}
 \caption{The logarithm of the vacuum energy density  between the plates for the free case as a function of $z/a$.}
 \label{fig0}
 \end{figure}

 The second term in  Eq. (\ref{T00free})  makes explicit the dependence of the energy density on the spatial coordinate $z$, perpendicular to 
 the plates. This term will be divergent on the plates for any dimension. This kind of divergences has been widely
 discussed in the literature \cite{plunien,mostepanenko,milton,Bordag,Bordag2,qftex09,lectphys}. There are different forms to deal with such divergences.
 One procedure to avoid surface divergences is by using the conformal stress tensor instead of the canonical one~\cite{coleman}. 
 Still using the minimal coupled stress-tensor one can avoid surface divergences treating the boundaries as quantum-mechanical objects~\cite{larry}. 
 In order to analyze  how this singular behavior near the boundaries is modified by the light-cone fluctuations we will present these local terms.
 To proceed with the case without randomness, we deal with such divergences through an analytic regularization procedure as follows.
 The local quantity $K(z)$ can be rewritten, by using the definition of the polylogarithm function
\begin{eqnarray}
 {\textrm{Li}}_{s}(z)=\sum_{n=1}^{\infty}\frac{z^n}{n^s},
\end{eqnarray}  
and its relation to the Hurwitz zeta function
\begin{eqnarray}
{\textrm{Li}}_{-s}(e^{\mu})=\frac{\Gamma(s+1)}{(2\pi)^{s+1}}\bigg[i^{s+1}\zeta\left(s+1,\frac{\mu}{2\pi i}\right)+
i^{-(s+1)}\zeta\left(s+1,1-\frac{\mu}{2\pi i }\right)\bigg],
\label{Li}
\end{eqnarray}
where $\zeta(s,a)$ is defined as
\begin{equation}
\zeta(s,a)=\sum_{n=0}^{\infty}(n+a)^{-s}, \hspace*{0.5cm} {\textrm{Re}}(s)>1, \hspace*{0.3cm} {\textrm{Re}}(a)>0,
\end{equation}
one can show that
    \begin{equation}
     K(z)=\frac{\Gamma(d+1)}{2(2\pi)^{d+1}}i^{d+1}\left(1+(-1)^{d+1}\right)\Bigl[\zeta(d+1,z/a)+\zeta(d+1,1-z/a)\Bigr].
     \label{k}
    \end{equation}
It can be seen that this local contribution does not vanish just for odd space dimensions. Accordingly, it is possible to write the free vacuum energy density as
\begin{equation}
 \langle T^{00}(z)\rangle_{0}= U_0(a) + g(z),
\end{equation}
where $g(z)$ is given by
\begin{equation}
g(z)=-\frac{(d-1)i^{d+1}}{(4a)^{d+1}}\frac{\Gamma(d+1)}{\pi^{1+d/2}}\Gamma\left(-\frac{d}{2}\right)
\Bigl[\zeta\left(d+1,z/a\right)+\zeta\left(d+1,1-z/a\right)\Bigr].
\end{equation}
\noindent To visualize the behavior of the renormalized vacuum energy density 
between the plates we analyze the case for $d=3$ spatial dimensions where we recover the known result~\cite{milton}
\begin{equation}\label{eq455}
 \langle T^{00}(z)\rangle_{0}= -\frac{\pi^2}{1440 a^4}-\frac{1}{16\pi^2 a^4}
 \Bigl[\zeta\left(4,z/a\right)+\zeta\left(4,1-z/a\right)\Bigr].
\end{equation}
The plot of this function is shown in Fig.~(\ref{fig0}), where the vacuum energy density is measured in units of $1/a^4$.

\noindent

It can be seen clearly the appearance of surface divergences on the vacuum energy density. It is worth to remark again that
the surface divergences would disappear once the Callan-Coleman-Jackiw conformal stress-energy tensor is employed instead of
the canonical stress tensor \cite{coleman}. Next, we can obtain the pressure in the $z$ direction on the plates from the equation
    \begin{equation}
    \langle T^{zz}(x)\rangle_{0}=\lim_{x'\rightarrow x}\frac{i}{2}\left(\partial_t\partial_{t'}-
    \partial_{\bf x_\bot}\partial_{\bf x'_\bot}+\partial_z\partial_{z'}\right)G_0(x,x').
    \end{equation}
 By taking the limit $x\rightarrow x'$ one notes that the pressure also depends on the confined coordinate $z$. It can be written as
    \begin{eqnarray}
    \langle T^{zz}(z)\rangle_{0}&=&\frac{i}{a}\int\frac{d\omega}{2\pi}\frac{d^{d-1}{\bf k}_\bot}{(2\pi)^{d-1}}
    \sum_{n=1}^\infty\bigg[\frac{\omega^2-\mathbf{k}_\bot^2}{\omega^2-\overline{\omega}_n^2+i\epsilon}\sin^2(n\pi z/a)\nonumber\\
    &&+\frac{(n\pi/a)^2}{\omega^2-\overline{\omega}_n^2+i\epsilon}\cos^2(n\pi z/a)\bigg].
    \end{eqnarray}
 Using an analytic regularization procedure and the reflection formula given by Eq.~(\ref{reflection}) one obtains the known result
    \begin{equation}\label{eqpress22}
    \langle T^{zz}\rangle_{0}=-\frac{d}{a^{d+1}}(4\pi)^{-(d+1)/2}\,\Gamma\left(\frac{d+1}{2}\right)\zeta(d+1),
    \end{equation}
 where the vacuum pressure of the free field is homogeneous inside the plates for all spatial dimensions~\cite{milton}. 
 In the next Section we will present the first-order contribution to the stress tensor due to randomness.


    \section{The renormalized vacuum expectation value of the stress tensor in the presence of light-cone fluctuations}\label{sec3}

\begin{figure}
\begin{center}
\includegraphics[width=0.9\textwidth]{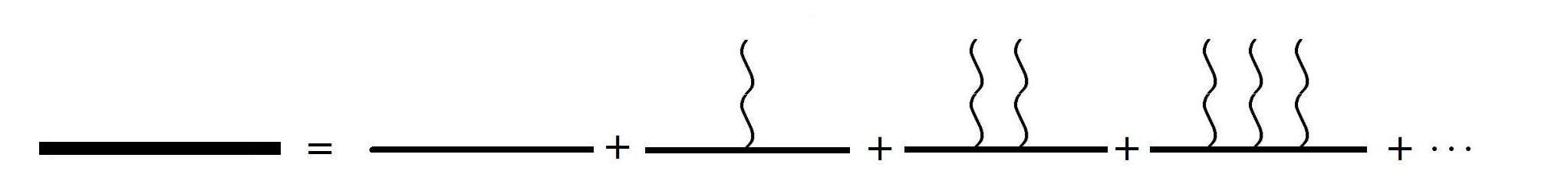}
\end{center}
\caption{Perturbative expansion of $G$ in terms of the disorder.
		   The wavy lines represent generically the random function $\nu$. }\label{expansion}
\end{figure}

The aim of this Section is to evaluate the stress tensor of the confined scalar field in the presence of light-cone fluctuations. 
As we discussed, in order to model such fluctuations we introduce randomness in the wave equation. We assume that Dirichlet boundary conditions 
can still be imposed on the scalar field. The random Klein-Gordon equation can be written as
\begin{eqnarray}
\left[\left(1 + g\nu(x) \right)\frac{\partial^{2}}{\partial\,t^{2}}-\nabla^{2}\right]
\varphi(x)=0,
\label{waveeq}
\end{eqnarray}
where $g$ is a small dimensionless parameter in order to implement a perturbative expansion. The random noise $\nu(x)$ has a Gaussian probability distribution defined by its moments
\begin{eqnarray}
\overline{\nu(x)}&=&0,\nonumber\\
\overline{\nu(x)\nu(x')}&=&\sigma^2\delta^{(d+1)}(x-x').
\end{eqnarray}
Here, the mean values over $\nu(x)$ has been depicted as $\overline{(...)}$ and $\sigma^2$ is the intensity of the noise. To implement a perturbative expansion we define the following characteristic length $l_c$ (in four dimensional space-time)
\begin{equation}
\frac{1}{l_c}=g^2\sigma^2\frac{\omega}{a^{4}}.
\label{lc}
\end{equation}
This expression arises from the self-energy induced by the interaction of the field with the random noise. Although we have chosen a space-time dependent noise, the white-noise type correlation allows us to define a steady characteristic length. It means that the space-like hypersurfaces associated with two different times are not correlated.
Therefore we can define a characteristic length independent of the time coordinate. Our perturbation theory will be valid for a given value of $g\ll 1$ such that it ensures that we are in the weak noise limit $l_c\ll a$. We can write the field equation as
\begin{equation}
(L_{0}+\,L_{1}) \, \varphi(x) = 0,
\end{equation}
where $L_0=\partial_t^2-\nabla^2$ is the usual differential operator and $L_1(x)=g\nu(x)\partial^2_{t}$ is a random differential operator. The full Green's function $G$ is given by
\begin{equation}
G=( L_{0} + L_{1} )^{-1}.
\end{equation}
In this limit, $G$ can be expanded in the following way
\begin{equation}
G = G_{0} - G_{0}\, L_1 \,G_{0} + G_{0}\, L_1 \,G_{0}\, L_1 \,G_{0}+ \, \cdots \nonumber \\
\end{equation}
$G_0=L_0^{-1}$ being the free Green's function. This perturbative expansion can be represented in a diagrammatic form shown in Fig.~(\ref{expansion}).
After performing the random averages, the first-order contribution to the Green's function due to the presence of random fluctuations is given by
    \begin{equation}
    G_{1}(x,x')=\int d^{d+1}x_1\,d^{d+1}x_2\,\overline{G_0(x,x_1) L_1(x_1) G_0(x_1,x_2) L_1(x_2)G_0(x_2,x')}.
    \end{equation}
From now on, we will include the parameter $g$ into the definition of the intensity of the noise $\sigma$. With all these considerations and using the Fourier representation given by Eq.~(\ref{G0}) we obtain
    \begin{eqnarray}
    G_{1}(x,x')&=&\sigma^2\int\frac{d\omega}{2\pi}\frac{d^{d-1}{\bf k}_\bot}{(2\pi)^{d-1}}
    \frac{d\omega'}{2\pi}\frac{d^{d-1}{\bf k'}_\bot}{(2\pi)^{d-1}}
    \exp[-i\omega(t-t')+i{\bf k}_\bot({\bf x}_\bot-{\bf x}'_\bot)]\nonumber\\
    &&\times\omega^2\omega'^2\int_0^a dz''{\cal G}_\lambda(z,z''){\cal G}_{\lambda'}(z'',z''){\cal G}_\lambda(z'',z').
    \end{eqnarray}
Here we can identify the self-energy induced by the metric fluctuations:
\begin{equation}
\Sigma(\omega;z)=\sigma^2\omega^2\int\frac{d\omega'}{2\pi}\frac{d^{d-1}{\bf k'}}{(2\pi)^{d-1}}\omega'^2{\cal G}_{\lambda'}(z,z).
\end{equation}
We denote the integral in the last equation by $H(z)$ and this is calculated in detail in the Appendix A. Accordingly Eq. (\ref{HAB}) the self-energy has the following form:
$\Sigma(\omega;z)\approx \sigma^2\omega^2/a^{4}$ this allows us to define the characteristic length by $\omega/l_c\approx\Sigma(\omega;z)$, as given by Eq.~(\ref{lc}).

Let us discuss on the implications of choosing a space-time dependent noise instead of a static noise. For a free massive scalar field with light-cone and mass static fluctuations, one has that the self-energy is proportional to $(\sigma^2_{\mu}\omega^4+\sigma^2_{\xi}m_0^4)|\omega^2-m_0^2|^{1/2}$ where $\sigma^2_{\mu}$ and $\sigma^2_{\xi}$ are the light-cone and mass noises intensities, respectively~\cite{arias}. In the limit of zero mass we define a characteristic length $l_c^{-1}\approx\sigma^2_{\mu}\omega^5$ which depends more sharply with frequency than the characteristic length for the case of a space-time dependent random fluctuations on the confined massless scalar field where we have $l_c^{-1}\approx\sigma^2\omega/a^4$. This is quite different for an unconfined massive scalar field in the presence of space-time dependent random fluctuations. In this case the characteristic length is proportional to the mass of the field $l_c^{-1}\approx\sigma^2\omega\,m^4$. Therefore in the zero mass limit the characteristic length is infinite and therefore the self-energy vanishes. This reveals the relevant role of the boundaries in order to define the weak-noise limit in the analysis of the massless scalar field with space-time dependent light-cone fluctuations.

Now let us calculate the first-order contribution to the stress tensor due to the randomness. As discussed in the previous Section, the renormalized vacuum energy density consists of two terms. One which is global and the other one is local. This behavior will persist in the presence of light-cone fluctuations. On the other hand, although the vacuum pressure of the free field is homogenous inside the plates, the corrections due to randomness introduces a local term similar to the one found for the energy density. The corrections to the vacuum energy density can be calculated from
    \begin{equation}
    \langle T^{00}(x)\rangle_{1}=\lim_{x'\rightarrow x}\frac{1}{2}\left(\partial_t\partial_{t'}+
    \partial_{\bf x_\bot}\partial_{\bf x'_\bot}+\partial_z\partial_{z'}\right)G_{1}(x,x').
    \end{equation}
In the limit $x\rightarrow x'$ we get
    \begin{eqnarray}
    \langle T^{00}(z)\rangle_{1}&=&\sigma^2\int\frac{d\omega}{(2\pi)}\frac{d^{d-1}{\bf k}_\bot}{(2\pi)^{d-1}}
    \omega^2\bigg[(\omega^2+\mathbf{k}_\bot^2)\int_0^a dz'\,{\cal G}_\lambda^2(z,z')H(z')\nonumber\\
    &&+\int_0^a dz'\,(\partial_z{\cal G}_\lambda(z,z'))^2H(z')\bigg],
    \label{26-energy}
     \end{eqnarray}
where the function $H(z)$ is given by
    \begin{equation}
    H(z)=\frac{2}{a}\sum_{n=1}^\infty\sin^2\left(\frac{n\pi z}{a}\right)
    \int\frac{d\omega}{2\pi}\frac{d^{d-1}{\bf k}_\bot}{(2\pi)^{d-1}}\frac{\omega^2}{\omega^2-\overline{\omega}_n^2-i\epsilon}.
\label{H27}    
\end{equation}
In order to obtain the above expression for $H(z)$, we have made use of Eq.~(\ref{gl0}). Note that $\langle T^{00}(z)\rangle_{1}$ depends on the confined coordinate $z$, similar to the free situation. Let us rewrite Eq.~(\ref{26-energy}) as
    \begin{equation}
    \langle T^{00}(z)\rangle_{1}=\sigma^2\int_0^a dz'\,H(z')\big(I(z,z')+J(z,z')\big),
    \label{28-energy}
\end{equation}
where the functions $I(z,z')$ and $J(z,z')$ are given respectively by
    \begin{eqnarray}
    I(z,z')&=&\frac{4}{a^2}\sum_{l,l'=1}^\infty
    \sin(l\pi z/a)\sin(l\pi z'/a)\sin(l'\pi z/a)\sin(l'\pi z'/a)\nonumber\\
    &&\times\int\frac{d\omega}{2\pi}\frac{d^{d-1}{\bf k}_\bot}{(2\pi)^{d-1}}
    \frac{\omega^2(\omega^2+\mathbf{k}_\bot^2)}{(\omega^2-\overline{\omega}_l^2+i\epsilon)(\omega^2-\overline{\omega}_{l'}^2+i\epsilon)},
    \nonumber\\
    J(z,z')&=&\frac{4}{a^2}\sum_{l,l'=1}^\infty
    \cos(l\pi z/a)\sin(l\pi z'/a)\cos(l'\pi z/a)\sin(l'\pi z'/a)\nonumber\\
    &&\!\!\!\times\left(\frac{ll'\pi^2}{a^2}\right)
    \int\frac{d\omega}{2\pi}\frac{d^{d-1}{\bf k}_\bot}{(2\pi)^{d-1}}
    \frac{\omega^2}{(\omega^2-\overline{\omega}_l^2+i\epsilon)(\omega^2-\overline{\omega}_{l'}^2+i\epsilon)}.
\label{IJ29}    
\end{eqnarray}
For other side the pressure in the $z$-direction inside the region confined by the plates can be calculated from the point-splitting formula
    \begin{equation}
    \langle T^{zz}(x)\rangle_{1}=\lim_{x'\rightarrow x}\frac{1}{2}\left(\partial_t\partial_{t'}-
    \partial_{\bf x_\bot}\partial_{\bf x'_\bot}+\partial_z\partial_{z'}\right)G_{1}(x,x').
    \end{equation}
Hence we get
    \begin{eqnarray}
    \langle T^{zz}(z)\rangle_{1}&=&\sigma^2\int\frac{d\omega}{(2\pi)}\frac{d^{d-1}{\bf k}_\bot}{(2\pi)^{d-1}}
    \omega^2\bigg[(\omega^2-\mathbf{k}_\bot^2)\int_0^a dz'{\cal G}_\lambda^2(z,z')H(z')\nonumber\\
    &&+\int_0^a dz'(\partial_z{\cal G}_\lambda(z,z'))^2H(z')\bigg].
    \end{eqnarray}
Again with the help of Eq.~(\ref{gl0}) we can rewrite the above expression as
    \begin{equation}
    \langle T^{zz}(z)\rangle_{1}=\sigma^2\int_0^a dz'\,H(z')\left[\tilde{I}(z,z')+J(z,z')\right],
\label{32-pressure}
\end{equation}
 where the function $\tilde{I}(z,z')$ is given by
    \begin{eqnarray}\label{eq322}
    \tilde{I}(z,z')&=&\frac{4}{a^2}\sum_{l,l'=1}^\infty
    \sin(l\pi z/a)\sin(l\pi z'/a)\sin(l'\pi z/a)\sin(l'\pi z'/a)\nonumber\\
    &&\times\int\frac{d\omega}{2\pi}\frac{d^{d-1}{\bf k}_\bot}{(2\pi)^{d-1}}
    \frac{\omega^2(\omega^2-\mathbf{k}_\bot^2)}{(\omega^2-\overline{\omega}_l^2+i\epsilon)(\omega^2-\overline{\omega}_{l'}^2+i\epsilon)}.
    \end{eqnarray}
 The stress-tensor components will be derived explicitly in \ref{ap1} and \ref{ap2}. In the next section we will use the results from the appendices to analyze the behavior of these components. Specially, we will focus on the local and global characteristics of the stress-tensor. As will be seen these terms will also present surface divergences, due to the local method used here, which need to be analyzed and properly renormalized.

\subsection{Local effects and surface divergences}   

\begin{figure}[t]
\begin{minipage}[b]{0.45\linewidth}
\centering
\includegraphics[width=\textwidth]{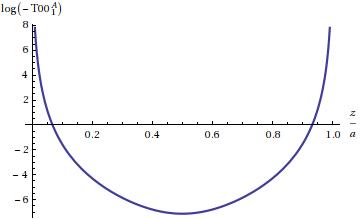}
\caption{The first correction, $\log(-\langle T^{00}(z)\rangle_{1}^A)$, as function $z/a$.}
\label{fig1}
\end{minipage}
\hspace{0.5cm}
\begin{minipage}[b]{0.45\linewidth}
\centering
\includegraphics[width=\textwidth]{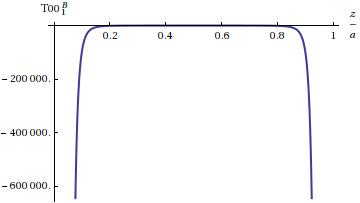}
\caption{The energy correction $\langle T^{00}(z)\rangle_{1}^B$ as function of $z/a$.}
\label{fig2}
\end{minipage}
\end{figure}

In this Section we present the  local effects  corrections due to light-cone fluctuations. Let us start our discussion with the vacuum energy density. In order to proceed, we need to regularize the contributions to the energy density. Using the results derived in the \ref{ap1}, we can decompose the energy density correction, Eq.~(\ref{28-energy}), as follows
    \begin{equation}
    \langle T^{00}(z)\rangle_{1}=\langle T^{00}(z)\rangle_{1}^A+\langle T^{00}(z)\rangle_{1}^B,
    \label{T001}
    \end{equation}
employing results from the appendices, in particular equations~(\ref{T00}),~(\ref{54-energy}) and~(\ref{TBB}), one have that
    \begin{eqnarray}\label{eq477}
    \langle T^{00}(z)\rangle_{1}^A&=&-\frac{\sigma^2}{a^{2d+2}}\frac{\zeta(d+1)}{(4\pi)^{d+1}}
    \left[\Gamma\left(\frac{d+1}{2}\right)\right]^2\Bigl[2\zeta(d+1)+(d-2)K(z)\Bigr],\nonumber\\       
       \langle T^{00}(z)\rangle_1^B&=&-\frac{\sigma^2}{a^{2d+2}2^{d+2}}\left(\frac{\pi}{4}\right)^d\left[\Gamma\left(-\frac{d}{2}\right)\right]^2
       \Bigl[U(z)-V(z)\Bigr],
      \end{eqnarray}
where the functions $U(z)$ and $V(z)$ are properly defined in \ref{ap2}. By using Eqs.~(\ref{k}),~(\ref{FG}) and~(\ref{xirho}) we can analyze the local behavior 
of the corrections to the vacuum energy density due to light-cone fluctuations. In Figs.~\ref{fig1} and~\ref{fig2} we present the local behavior of these corrections 
between the plates. There the corrections to the vacuum energy density are measured in units of $\sigma^2/a^8$.
To implement the correction due the randonmess we made use of the perturbation theory presented in Section \ref{sec3}. This procedure must be valid in the whole region between the plates. However as we can see by Fig. \ref{fig2}, the correction become larger when close to the boundaries. As this effect is multiplied by $\sigma^2/a^8$ a good estimative to the $\sigma$-factor is that if one measure the energy density in a given point in the region between the plates and close to one of them, the final result must be much smaller than the case without fluctuations, discussed in Section \ref{sec2}.  Putting this in numerical values, if the energy density is measured in a point $z = 0.05a$, with $a$ the distance between the plates, the zero-order term to the energy density is $\langle T^{00}\rangle_0 \approx 10^3/a^4$ while the correction $|\langle T^{00}\rangle_1^A + \langle T^{00}\rangle_1^B| \approx 10^7\sigma^2/a^8$. Then, a possible estimative is $\sigma^2 \ll 10^{-4}a^4$.

The Figs. \ref{fig1} and \ref{fig2} also show that the general feature of surface divergences is still the same as in the free case. As discussed before, for the free case the surface divergences could be avoided by using the conformal stress-tensor \cite{coleman}. However the general belief that surface divergences will disappear once the conformal stress-tensor is used instead of the canonical one, is not a general fact. It has been shown that once we consider curved boundaries instead of flat ones, the surface divergences arise even if we are using the conformal stress-tensor \cite{Green's3}.In another related work it was shown that even with flat boundaries, when mixed boundaries conditions are considered the surface divergences also could remain for the conformal stress-tensor \cite{romeo}. We conclude that the use of the improved conformal stress-tensor does not allow us to remove surface divergences in the above discussed cases. For the case of flat boundaries and Dirichlet boundary conditions in the presence of light-cone fluctuations, the surface divergences arises from the functions $K(z)$, $U(z)$ and $V(z)$. These local terms with surface divergences could not be completely removed by using the conformal stress-tensor. On the other hand, it can be seen that the stochastic light-cone fluctuations induces a self-interaction in the original free scalar field. This self-interaction will be qualitatively similar to the $\lambda\varphi^4$ theory. In general, it is well known that in self-interacting non-translational invariant systems  to render the theory finite is necessary to introduce not only bulk counter-terms but also surface counter-terms \cite{symansik, fosco, caicedo, nami5, martin}. In conclusion, to avoid surface divergences, it is necessary to introduce singular counterterms in order to eliminate a singular surface energy density. This procedure will not be explicitly performed here, instead  we refer the reader to an earlier work \cite{martin} and we will concentrate our discussion on the global physical terms which are free of ambiguities. 

We can apply the same reasoning to the vacuum pressure. Note that the vacuum pressure as well as the vacuum energy density now depends on the distance to the plates, as mentioned above. This situation is different from the free case. Following a similar procedure as the one considered for the energy density, we get
    \begin{equation}
    \langle T^{zz}(z)\rangle_{1}=\langle T^{zz}(z)\rangle_{1}^A+\langle T^{zz}(z)\rangle_{1}^B,
    \label{Tzz}
    \end{equation}
where using the results derived in the \ref{ap1} we have
    \begin{eqnarray}\label{eqtzz1}
    \langle T^{zz}(z)\rangle_{1}^A&=&-\frac{\sigma^2}{4a^{2d+2}}\left(\frac{\pi}{4}\right)^{d}
    \left[\Gamma\left(-\frac{d}{2}\right)\right]^2\zeta(-d)
    \bigg[(d+1)\zeta(-d)-K(z)\bigg],\nonumber\\       
       \langle T^{zz}(z)\rangle_{1}^B&=&\frac{\sigma^2{\cal N}_3}{a^{d+2}}\sum_{l,l'=1}^\infty\,M_B(l,l')\tilde{R}_{l,\,l'}(z).
      \end{eqnarray}
Both terms show a similar behavior in comparison with Figs.~(\ref{fig1}) and~(\ref{fig2}). In particular we have a important difference from the case without light-cone fluctuations, 
namely the fact that light-cone fluctuations introduce a vacuum pressure correction dependent on the distance to the plates. However, as discussed before, this behavior will not be considered. In the next Section we will present the global terms that give rise to corrections to the Casimir effect due to random light-cone fluctuations.

       \subsection{Global Casimir terms}
 
Now let us focus on the (physical) global terms that will contribute to the Casimir effect. Avoiding the local terms in Eq. (\ref{eq477}) as discussed previously, we have that the contribution to the energy density per unit area of the plates is given by
    \begin{equation}\label{eq35}
    u_1=-\frac{2\sigma^2}{a^{2d+1}(4\pi)^{d+1}}
    \bigg[\Gamma\bigg(\frac{d+1}{2}\bigg)\zeta(d+1)\bigg]^2.
    \end{equation}
For more details of such a computation we refer the reader to the \ref{ap1}. Note that in the case $d=3$, the energy density per unit area is proportional to $a^{-7}$, while energy density per volume is proportional to $a^{-8}$ (see Eq. (\ref{eq477})). This result resembles the effects of Van der Waals interactions~\cite{milton,plunien,mostepanenko}. However, they are not the same effect, once the $a^{-7}$ behavior present in the perturbed term shown here is only due the random parameter. So this coincidence seems accidental.

Also the increase of the vacuum pressure on the plates due to light-cone fluctuations can be obtained. Denoting this correction as $p_1$, this contribution will be given by the global term in $\langle T^{zz}(z)\rangle_{1}$ (see Eq. (\ref{eqtzz1})). So, we have that the contributions of the randomness for the vacuum pressure on each plate are given by
    \begin{equation}
    p_1=-\frac{\sigma^2\,(d+1)}{a^{2d+2}(4\pi)^{d+1}}
   \bigg[\Gamma\bigg(\frac{d+1}{2}\bigg)\zeta(d+1)\bigg]^2.
    \end{equation}
As can be seen from the expression above $p_1$ is clearly negative. Thus light-cone fluctuations in the 
weak-noise limit induce a small increase in the attraction between the plates.


    \section{Conclusions}\label{sec5}

The aim of this paper was to investigate the renormalized stress-tensor associated with the scalar field obeying a wave equation with a space-time dependent random coefficient. We assumed that the quantum field is constrained by the presence of material boundaries. Thermal Casimir effect between random layered dielectrics was studied in Ref.~\cite{di4}. 
In turn, a scalar wave equation where the dielectric constant has a time-dependent random contribution was discussed by Stephen~\cite{time}. As mentioned above we proposed a situation where both conditions are present. We have found that the corrections to the energy density between the plates in four space-time dimensions goes as $1/a^{8}$. 
We remark that the situation considered here is a simplified model for the more realistic case of the Casimir energy due to nonrelativistic phonons in a disordered fluid confined between plane boundaries. 

One of the main results of this work is the fact that due to the fluctuations, the renormalized correction for the vacuum pressure depends on the distance to the plates in a non-trivial way, namely it shows a dependence on the confined coordinate $z$. Such a dependence does not appears in the case without light-cone fluctuations. This indicates that random noise breaks down the pressure homogeneity between the plates. The local behavior of the vacuum energy density and vacuum pressure is analogous to the local effects encountered in the non-disorder Casimir effect. We emphasize that the attractive character of the Casimir force for the geometry considered here still remains and the random corrections increase its intensity, even though by a small amount. 

Another important result concerns the surface divergences present in $\langle T^{00}(z)\rangle_1$ and $\langle T^{zz}(z)\rangle_1$. It is well known in the literature that such divergences are usually eliminated when one considers the conformal stress-tensor instead of the canonical one. This is what happen in the free case presented in Sec. (\ref{sec3}), for instance. However, for the random case this procedure does not remove the divergences in $z = 0, a$, see Eqs. (\ref{eq477}) and (\ref{eqtzz1}). This result is  similar to the result found in Ref. \cite{Green's3}, where for a curved boundary the surface divergences do not disappears.

A natural extension of this paper is to investigate the Casimir energy due to non-relativistic phonons 
in a disordered fluid confined between flat boundaries. This subject is under investigation by the authors.


    \section{Acknowledgments}

    This work was partially financed by CNPq and FAPERJ (Brazilian agencies). The authors would like to thanks L. H. Ford and V. A. de Lorenci for helpful discussions.


    \appendix
    \section{Corrections to the stress-energy tensor induced by the random fluctuations}\label{ap1}
   
In this Appendix we will show in detail the procedure employed to obtain the corrections to the stress-energy tensor induced by the light-cone fluctuations discussed in the text. We will start with the expression for the function $H(y)$ given by Eq.~(\ref{H27}). The integral over frequencies in such an expression can be performed in the usual way. Performing also the integral over the transverse momenta we get
    \begin{eqnarray}
    H(z)&=&\frac{2{\cal N}_1}{a^{d+1}}\sum_{n=1}^\infty \sin^2\left(\frac{n\pi z}{a}\right)n^d,\nonumber\\
    &=&\frac{{\cal N}_1}{a^{d+1}}\sum_{n=1}^\infty \left[1-\cos\left(\frac{2n\pi z}{a}\right)\right]n^d,\nonumber\\
    &=&H^A(z)+H^B(z),
    \end{eqnarray}
where we have defined
    \begin{eqnarray}
    H^A(z)&=&\frac{{\cal N}_1}{a^{d+1}}\zeta(-d),\nonumber\\
    H^B(z)&=&-\frac{{\cal N}_1}{a^{d+1}}\,K(z),
    \label{HAB}
    \end{eqnarray}
 and the constant ${\cal N}_1$ is defined by
    \begin{equation}\label{eq499}
    {\cal N}_1=\frac{i}{2}\left(\frac{\pi}{4}\right)^{d/2}\Gamma\left(-\frac{d}{2}\right).
    \end{equation}
The function $K(y)$ was defined previously [see Eqs.~(\ref{k12}) and~(\ref{k})]. In order to present a regularized expression for the renormalized vacuum energy density, one needs to consider the functions $I(y,z)$ and $J(y,z)$ given by Eq.~(\ref{IJ29}). Performing the $\omega$-integrals using complex variables and integrating over the transverse momenta \cite{erdelyi,abram}, we obtain
    \begin{eqnarray}\label{eq54}
    I(z,z')&=&\frac{{\cal N}_2}{a^{d+2}}\sum_{l,l'=1}^\infty
    \sin(l\pi z/a)\sin(l\pi z'/a)\sin(l'\pi z/a)\sin(l'\pi z'/a)\nonumber\\
    &&\,\,\,\,\,\,\,\,\,\,\,\,\,\,\,\,\,\,\,\,\,\,\,\times\frac{l^{d+2}-l'^{d+2}}{l^2-l'^2},
    \nonumber\\
    J(z,z')&=&\frac{{\cal N}_3}{a^{d+2}}\sum_{l,l'=1}^\infty
    \cos(l\pi z'/a)\sin(l\pi z/a)\cos(l'\pi z'/a)\sin(l'\pi z/a)\nonumber\\
    &&\,\,\,\,\,\,\,\,\,\,\,\,\,\,\,\,\,\,\,\,\,\,\,\times
    (l\,l')\frac{l^{d}-l'^{d}}{l^2-l'^2},\nonumber\\
    \end{eqnarray}
 where
    \begin{eqnarray}\label{eq55}
    {\cal N}_2&=&-2i\left(\frac{\pi}{4}\right)^{d/2}
    \left(\frac{d-4}{d+2}\right)\Gamma\left(-\frac{d}{2}\right),\nonumber\\
    {\cal N}_3&=&2i\left(\frac{\pi}{4}\right)^{d/2}\Gamma\left(-\frac{d}{2}\right).
    \end{eqnarray}
 Note that, ${\cal N}_2=(4-d){\cal N}_3/(2+d).$
 From Eqs. (\ref{eq54}) and~(\ref{eq55}) we find
    \begin{equation}
    I(z,z')+J(z,z')=\frac{{\cal N}_3}{a^{d+2}}\sum_{l,l'=1}^\infty\sin(l\pi z/a)\sin(l'\pi z/a)R_{l,\,l'}(z'),
    \end{equation}
 where we have defined
    \begin{eqnarray}\label{eq555}
    R_{l,\,l'}(z')\!&=&\!\bigg[\frac{4-d}{2+d}\sin(l\pi z'/a)\sin(l'\pi z'/a)\left(\frac{l^{d+2}-l'^{d+2}}{l^2-l'^2}\right)\nonumber\\
    &&{}\qquad+\cos(l\pi z'/a)\cos(l'\pi z'/a)\,ll'\left(\frac{l^{d}-l'^{d}}{l^2-l'^2}\right)\bigg].
    \label{Rll}
    \end{eqnarray}
 Thus from Eq.~(\ref{28-energy}) the contribution to the vacuum energy density is given by
    \begin{eqnarray}
    \langle T^{00}(z)\rangle_{1}&=&\sigma^2\int_0^a dz'\,H(z')\left[I(z,z')+J(z,z')\right]\nonumber\\
    &=&\frac{\sigma^2{\cal N}_3}{a^{d+2}}\sum_{l,l'=1}^\infty M(l,l')\,R_{l,\,l'}(z).
		\end{eqnarray}
 The quantity $M(l,l')$ can be written down as the sum of two terms
    \begin{eqnarray}
    M(l,l')&=&
    \int_0^a\,dz'H(z')\sin(l\pi z'/a)\sin(l'\pi z'/a),\nonumber\\
    &=&M_A(l,l')+M_B(l,l').
    \end{eqnarray}
 These are defined with the help of Eq.~(\ref{HAB}) as
    \begin{eqnarray}
    M_A(l,l')&=&\frac{{\cal N}_1\,\zeta(-d)}{a^{d+1}}	\int_0^a\,dz'\sin(l\pi z'/a)\sin(l'\pi z'/a),\nonumber\\
    M_B(l,l')&=&-\frac{{\cal N}_1}{a^{d+1}}\int_0^a\,dz'\,K(z')\sin(l\pi z'/a)\sin(l'\pi z'/a).
    \label{MAB}
    \end{eqnarray}
Therefore the following decomposition can be held
    \begin{equation}
    \langle T^{00}(z)\rangle_{1}=\langle T^{00}(z)\rangle_{1}^A+\langle T^{00}(z)\rangle_{1}^B,
    \label{T00}
    \end{equation}
 where
    \begin{eqnarray}
    \langle T^{00}(z)\rangle_{1}^A&=&\frac{\sigma^2{\cal N}_3}{a^{d+2}}\sum_{l,l'=1}^\infty M_A(l,l')\,R_{l,\,l'}(z),\nonumber\\
    \langle T^{00}(z)\rangle_{1}^B&=&\frac{\sigma^2{\cal N}_3}{a^{d+2}}\sum_{l,l'=1}^\infty M_B(l,l')\,R_{l,\,l'}(z).
    \label{TAB}
    \end{eqnarray}
Let us focus on the first term on the right-hand side of Eq.~(\ref{T00}). The second term has a more cumbersome expression, which we will work out in detail in the next Appendix. Due to the orthogonality of sine functions $$\int_0^\pi d\alpha\,\sin(l\alpha)\sin(l'\alpha)=(\pi/2)\delta_{l,\,l'},$$	we find that
    \begin{equation}
    M_A(l,l')=\frac{{\cal N}_1}{2a^d}\zeta(-d)\delta_{l,\,l'}.
    \end{equation}
Therefore the sum over $l'$ in the first expression of Eq.~(\ref{TAB}) can be easily performed. Considering the limit $l'\rightarrow l$ in Eq.~(\ref{Rll}) yields us terms proportional to $l^{d}$. Hence
    \begin{eqnarray}
    \langle T^{00}(z)\rangle_{1}^A &=&\frac{\sigma^2\zeta(-d)}{4a^{2d+2}}{\cal N}_1{\cal N}_3
    \sum_{l=1}^\infty\,l^{d}(2+(d-2)\cos(2l\pi z/a)),\nonumber\\
    &=&\frac{\sigma^2 \zeta(-d)}{4a^{2d+2}}{\cal N}_1{\cal N}_3[2\zeta(-d)+ (d-2)K(z)].	
    \label{54-energy}    
     \end{eqnarray}
 Integrating in the bulk region between the plates, we obtain the first contribution to the vacuum energy density per unit area of the plates due to light-cone fluctuations\\ $u_1=\int_0^a dz\langle T^{00}(z)\rangle_{1}$. Similar to the free case, the function $K(z)$ will bring surface divergences that can be avoid considering the conformal stress-tensor.     
Taking into account only the global term Eq. (\ref{54-energy}) we get, after replacing the expressions for ${\cal{N}}_1$ and ${\cal{N}}_3$
    \begin{eqnarray}
    u_1^{A}&=&-\frac{2\sigma^2}{a^{2d+1}(4\pi)^{d+1}}
    \bigg[\Gamma\bigg(\frac{d+1}{2}\bigg)\zeta(d+1)\bigg]^2.
    \end{eqnarray}
 Even though $\langle T^{00}(z)\rangle_{1}^B$ has a more involved expression than $\langle T^{00}(z)\rangle_{1}^A$ as discussed above, through a straightforward calculation 
 we can present its contribution to the vacuum energy density per unit area of the plates. Due to the orthogonality of sine functions given above we first integrate in the bulk (cosine functions has an analogous relation). We get
    \begin{eqnarray}\label{eq622}
    u_1^B&=&\int_0^a\,dz \langle T^{00}(z)\rangle_{1}^B=
    \frac{\sigma^2{\cal N}_3}{a^{d+2}}\sum_{l,l'=1}^\infty M_B(l,l')\,\int_0^a\,dzR_{l,\,l'}(z).
    \end{eqnarray}
 After integration over $z$ one may perform the sum over $l'$. Again taking the limit $l\rightarrow l'$ in $R_{l,\,l'}$, we get terms proportional to $l^{d}$. Therefore $$\sum_{l'}M_B(l,l')R_{l,\,l'}=al^{d}M_B(l,l).$$ Remembering the definition of the $K(z)$ function in Eq.~(\ref{k12}), one has
    \begin{eqnarray}
        M_B(l,l)&=& -\frac{{\cal N}_1}{a^{d+1}}\sum_{n=1}^\infty n^d
         \int_0^a\,dz' \cos(2\pi nz'/a)\sin^2(l\pi z'/a),\nonumber\\
         &=& -\frac{{\cal N}_1}{2a^{d+1}}\sum_{n=1}^\infty n^d
         \int_0^a\,dz' \cos(2\pi nz'/a)\bigg(1-\cos(2l\pi z'/a)\bigg),\nonumber\\
         &=& \frac{{\cal N}_1}{4a^d} l^{d}.
    \end{eqnarray}
Inserting this last result in Eq.~(\ref{eq622}) leads us to
    \begin{equation}
    u_1^B=-\frac{\sigma^2}{4a^{2d+1}}\bigg(\frac{\pi}{4}\bigg)^{d}\bigg[\Gamma\bigg(-\frac{d}{2}\bigg)\bigg]^2\zeta(-2d),
    \end{equation}
where Eqs.~(\ref{eq499}) and~(\ref{eq55}) were used. We can see that this contribution $u_1^B$ is ill defined for even spatial dimensions $d$. For odd spatial dimensions this contributions is just null. We will see in the next appendix that this contributions comes from a purely local term in the energy component of the stress-tensor. This kind of terms are present due to the non-univocality of the stress-tensor. By using the conformal stress-tensor some of these terms could be avoided but in general surface counter-terms have to included in order to render the theory finite.

		Now following a similar procedure we will evaluate the vacuum pressure. As above we have the following decomposition
    \begin{eqnarray}
    \langle T^{zz}(z)\rangle_{1} &=& \sigma^2\int_0^adz'\, H(z')(\tilde{I}(z,z')+J(z,z')),\nonumber\\
    &=&\langle T^{zz}(z)\rangle_{1}^A+\langle T^{zz}(z)\rangle_{1}^B,
    \end{eqnarray}
 where $\tilde{I}(z,z')$ and $J(z,z')$ are given by Eqs.~(\ref{eq322}) and~(\ref{eq54}), respectively. Then $\langle T^{zz}(z)\rangle_{1}^A$ is given by
    \begin{equation}
    \langle T^{zz}(z)\rangle_{1}^A=\frac{\sigma^2}{a^{d+2}}{\cal N}_3\sum_{l,l'=1}^\infty\,M_A(l,l')\tilde{R}_{l,\,l'}(z),
    \end{equation}
 where we have defined
    \begin{eqnarray}
    \tilde{R}_{l,\,l'}(z)\!&=&\!\bigg[\sin(l\pi z/a)\sin(l'\pi z/a)\left(\frac{l^{d+2}-l'^{d+2}}{l^2-l'^2}\right)\nonumber\\
    &&{}\qquad+\cos(l\pi z/a)\cos(l'\pi z/a)\,ll'\left(\frac{l^{d}-l'^{d}}{l^2-l'^2}\right)\bigg].
    \end{eqnarray}
 Following similar steps it can be shown that
    \begin{equation}\label{eq699}
    \langle T^{zz}(z)\rangle_{1}^A=-\frac{\sigma^2}{4a^{2d+2}}\left(\frac{\pi}{4}\right)^{d}
    \left[\Gamma\left(-\frac{d}{2}\right)\right]^2\zeta(-d)
    \bigg[(d+1)\zeta(-d)-K(z)\bigg].
    \end{equation}
The other pressure term is given by
    \begin{equation}\label{eq699B}
    \langle T^{zz}(z)\rangle_{1}^B=\frac{\sigma^2{\cal N}_3}{a^{d+2}}\sum_{l,l'=1}^\infty\,M_B(l,l')\tilde{R}_{l,\,l'}(z).
    \end{equation}
As mentioned in the text, both terms show a similar behavior in comparison with Figs.~(\ref{fig1}) and~(\ref{fig2}). 

    \section{Calculation of the second contribution to the energy density $\langle T^{00}(z)\rangle_1^B$}\label{ap2}
In this appendix we perform the calculation of the second contribution to the vacuum energy density $\langle T^{00}(z)\rangle_1^B$ due to light-cone fluctuations. In order to perform the integral in $M_B(l,l')$ we will use the relation $$\sin(l\pi z/a)\sin(l'\pi z/a)=\frac{1}{2}(\cos(|l-l'|\pi z/a)-\cos((l+l')\pi z/a)),$$ and the orthogonality property of the cosine functions $\cos(n\theta)$. We obtain
		\begin{equation}
    M_B(l,l')=\frac{{\cal N}_1}{4a^{d}}\sum_{n}n^d\,(\delta_{2n,l+l'}-\delta_{2n,|l-l'|}).
    \label{z111}
    \end{equation}
In Eq. (\ref{z111}) we have non-null terms only when $2n=|l\pm l'|$. This will restrict the values of $l$ and $l'$ over which the sum in the second expression of Eq.~(\ref{TAB}) can be performed. Since $n$ is an integer number, the result of $l\pm l'$  must be an even number. Therefore $l$ e $l'$ should have the same parity. Inserting Eq.~(\ref{z111}) in the expression for $ \langle T^{00}(z)\rangle_1^B$ yields
     \begin{equation}
    \langle T^{00}(z)\rangle_1^B=\frac{\sigma^2 {\cal N}_3{\cal N}_1}{a^{2d+2}2^{d+2}}
    \left[U(z)-V(z)\right],
    \label{TBB}
    \end{equation}
 where we have defined the functions
    \begin{eqnarray}
    U(z)&=&\sum^{*}_{l,l'}(l+l')^dR_{l,l'}(z),\nonumber\\
    V(z)&=&\sum^{*}_{l,l'}|l-l'|^dR_{l,l'}(z).
    \label{UV}
    \end{eqnarray}
 The symbol $*$ in Eq.~(\ref{UV}) means that the sum is restricted to values of $l$ and $l'$ such that they have the same parity, namely
    \begin{equation}
    \sum^{*}_{l,l'}=\sum_{l=2m,\, l'=2m'}+\sum_{l=2m+1,\, l'=2m'+1}
    \label{*}
    \end{equation}
 By using Eq.~(\ref{*}) in~(\ref{UV}) we obtain
    \begin{equation}
    U(z)=U_I(z)+U_{II}(z),
    \label{U}
    \end{equation}
 where
    \begin{eqnarray}
    U_I(z)&=&2^d\sum_{m,m'}(m+m')^dR_{2m,2m'}(z),\nonumber\\
    U_{II}(z)&=&2^d\sum_{m,m'}(m+m'+1)^dR_{2m+1,2m'+1}(z).
    \label{UI}
    \end{eqnarray}
 Remembering the definition of the function $R_{l,\,l'}(z)$ given by Eq.~(\ref{eq555}) one observes that
    \begin{equation}
    R_{2m,2m'}(z)=2^dR_{m,m'}(2z).
    \end{equation}
 Hence, the first contribution in Eq. (\ref{UI}) could be written as
    \begin{equation}
    U_I(z)=2^{2d}\sum_{m,m'}(m+m')^d\,R_{m,m'}(2z).
    \label{Rmm}
    \end{equation}
In Eq. (\ref{Rmm}) the sum in $m$ and $m'$ have no restriction. In this equation, the Newton's generalized binomial theorem can be employed. In addition the denominators coming from the function $R_{m,m'}(2z)$ can be expanded as a sum of products of powers of $m$ and $m'$. This enables one to write the double sums in $l,l'$ as products of two terms, namely $\sum_{l,l'}f(l, l') = (\sum_l f(l))(\sum_{l'}f(l'))$. One obtains
    \begin{eqnarray}
    U_I(z)&=&
    2^{2d}\sum_{k=0}^{d-1}\frac{(d-1)!}{k!(d-1-k)!}\nonumber\\&&
    \times\bigg[\frac{4-d}{2+d}\sum_{i=1}^{d+2}{\cal F}(2d-k-i+1;2z){\cal F}(k+i-1;2z)\nonumber\\
    &&+\sum_{j=1}^{d}{\cal G}(2d-k-j;2z){\cal G}(k+j;2z)\bigg].     
    \label{TBfg}
    \end{eqnarray}
In Eq. (\ref{TBfg}) we have defined the functions 
    \begin{eqnarray}
    {\cal F}(m;z)&=&\sum_l\,l^m\sin(l\pi z/a),\nonumber\\
    {\cal G}(m;z)&=&\sum_l\,l^m\cos(l\pi z/a).
    \end{eqnarray}
By using the properties of the polylogarithm function and the Hurwitz Zeta function we get
    \begin{eqnarray}
    {\cal F}(m;z)&=&\frac{\Gamma(m+1)}{2(2\pi)^{m+1}}i^{m}(1+(-1)^m)\left[\zeta(m+1,z/2a)+\zeta(m+1,1-z/2a)\right],\nonumber\\
    {\cal G}(m;z)&=&\frac{\Gamma(m+1)}{2(2\pi)^{m+1}}i^{m+1}(1+(-1)^{m+1})\left[\zeta(m+1,z/2a)-\zeta(m+1,1-z/2a)\right].\nonumber\\
    \label{FG}
    \end{eqnarray}
From Eq. (\ref{FG}) we see that ${\cal F}(2n+1;z)={\cal G}(2n;z)=0$ for any integer number $n$. Therefore in Eq. (\ref{TBfg}) we have just a few non-null terms. For the case $d=3$, we have
    \begin{eqnarray}
    U_I(z)&=&\frac{128}{5}\bigg(4{\cal F}(2,2z){\cal F}(4,2z)
    +{\cal F}(0,2z){\cal F}(6,2z)\nonumber\\&&+10({\cal G}(3,2z))^2+5{\cal G}(1,2z){\cal G}(5,2z)
    \bigg).
    \end{eqnarray}
Now let us focus on the other contribution to the function $U(z)$. From Eq.~(\ref{UI}) we see that
    \begin{eqnarray}
    U_{II}(z)=2^d\sum_{m,m'}(m+m'+1)^dR_{2m+1,2m'+1}(z).
    \label{}
    \end{eqnarray}
Following an analogous procedure as discussed above, for $d=3$ we get
    \begin{eqnarray}
    U_{II}(z)&=&2^3\sum_{j=0}^{2}\sum_{k''=0}^{j}C^{2}_jC^{j}_{k''}\nonumber\\
    &&\times\bigg\{\frac{1}{5}\sum_{i=1}^{5}\sum_{k=0}^{5-i}\sum_{k'=0}^{i-1}C^{5-i}_kC^{i-1}_{k'}2^{k+k'-1}\xi(j+k-k'';z)\xi(k'+k'';z)
    \nonumber\\
   &&+\sum_{i=1}^{3}\sum_{k=0}^{3-i}\sum_{k'=0}^{i-1}C^{3-i}_kC^{i-1}_{k'}2^{k+k'-1}\bigg[4\rho(j-k-k''+1;z)\rho(k'+k''+1;z)\nonumber\\
    &&+2\rho(j-k+k''+1;z)\rho(k'+k'';z)+2\rho(j-k+k'';z)\rho(k'+k''+1;z)\nonumber\\
    &&+\rho(j-k+k'';z)\rho(k'+k'';z)\bigg]\bigg\}.
    \end{eqnarray}
where $C^n_k$ are combinatorial factors $$C^n_k=\frac{n!}{k!(n-k)!},$$ and we have defined the functions 
    \begin{eqnarray}
    \xi(n,z)&=&\sum_m m^n\sin((2m+1)\pi z/a),\nonumber\\
    \rho(n,z)&=&\sum_m m^n\cos((2m+1)\pi z/a).
    \end{eqnarray}
Performing the above sums we get 
    \begin{eqnarray}
    \xi(n,z)&=&{\cal G}(n,2z)\sin(\pi z/a)+{\cal F}(n,2z)\cos(\pi z/a),\nonumber\\
    \rho(n,z)&=&{\cal G}(n,2z)\cos(\pi z/a)-{\cal F}(n,2z)\sin(\pi z/a).
    \label{xirho}
    \end{eqnarray}
 By using the same reasoning, it can be shown that the function $V(z)$ appearing in Eq.~(\ref{UV}) exhibits a similar behavior as a function of $z$ in comparison with $U(z)$.

\end{document}